\documentclass[aps,reprint,nobibnotes,nofootinbib,superscriptaddress]{revtex4-2}

\usepackage{graphicx, amstext,url,verbatim} 
\usepackage{mathtools,array,xcolor,physics}
\usepackage[per-mode=symbol]{siunitx}
\graphicspath{ {./Final/} }

\begin{document}
\title{Free Growth under Tension}
\date{February 5, 2025}
\author{Chenyun Yao}
\author{Jens Elgeti}
\email{j.elgeti@fz-juelich.de}
\affiliation{Theoretical Physics of Living Matter, Institute for Advanced Simulation, Forschungszentrum Jülich, Jülich, Germany}

\begin{abstract}
	Ever since the ground breaking work of Trepat et al. in 2009, we know that cell colonies growing on a substrate can be under tensile mechanical stress.
	The origin of tension has so far been attributed to cellular motility forces being oriented outward of the colony.
	Works in the field mainly revolve around how this orientation of the forces can be explained, ranging from velocity alignment, self-sorting due to self-propulsion, to kenotaxis.

  In this work, we demonstrate that tension in growing colonies can also be explained \textit{without} cellular motility forces!
  Using a combination of well established tissue growth simulation technique and analytical modelling, we show how tension can arise as a consequence of simple mechanics of growing tissues.
	Combining these models with a minimalistic motility model shows how colonies can expand while under even larger tension.
  Furthermore, our results and analytical models provide novel analysis procedures to identify the underlying mechanics. 
\end{abstract}
\maketitle

\begin{figure}
  \includegraphics[width=\columnwidth]{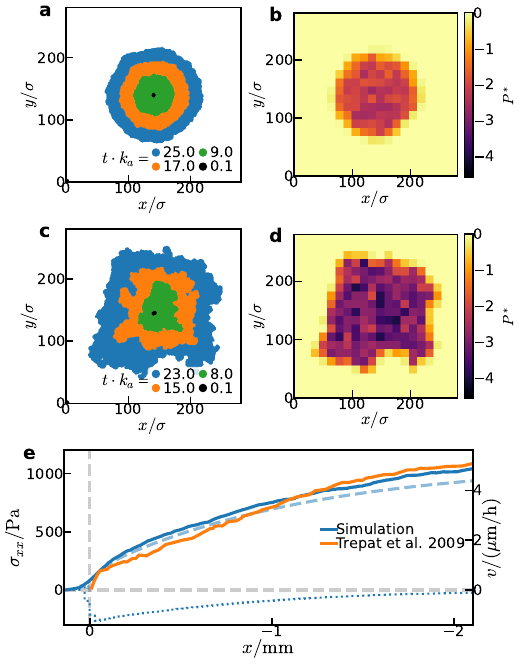}
	\caption{ Growth of tensile colonies.
	(a) Superimposed snapshots of a growing colony of non-motile cells~($\varepsilon^*=2, G^*=16.4, v_0^*=0$) at various times.
	It grows indefinitely in a roughly circular shape, but remains under tension.
	(b) The local stress profile corresponding to the last snapshot~(blue, $t\cdot k_a=25$) of (a).
	(c) Superimposed snapshots of a growing colony of motile cells~($\varepsilon^*=2, G^*=15.0, v_0^*=42.1$) at various times.
	It grows indefinitely, displaying fingering at the edge and large tension inside.
	(d) The local stress profile corresponding to the last snapshot~(blue, $t\cdot k_a=23$) of (c). 
	Both colonies are under tension, which is built up from the boundary to the center.
	(e) The stress profile of an expanding quais-1d motilie colony~(blue solid) with $\varepsilon^*=2, G^*=17.8, v_0^*=238.5 $ matches those from Trepat et al. 2009~\cite{trepat_physical_2009} by taking $\sigma=20\si{\micro\meter}, k_a^{-1}=120\si{\hour}, \varepsilon=2\times 10^{-12}\si{\joule}.$
		With such choice of parameters, the motility is $40\si{\micro\meter\per\hour}$, the max retrograde flow is $1\si{\micro\meter\per\hour}$, the expansion speed is $0.1\si{\micro\meter\per\hour}$.
		The dashed and dotted lines are the corresponding theory prediction and veloicty profile.
	}
	\label{fig:1}
\end{figure}

Just as biochemical conditions, mechanics can affect the growth of biological tissues~\cite{montel_stress_2011,delarue_mechanical_2013,hallatschek_proliferating_2023,shraiman_mechanical_2005,belteton_real-time_2021}.
As the conjugate force to cell volume, particular attention has been given to pressure or tension. 
When a tissue grows, it exerts forces on its surroundings
and vice versa experiences the reaction force.
It is generally assumed~\cite{basan_homeostatic_2009} and experimantally confirmed~\cite{montel_stress_2011}, that pressure reduces growth.
The idea is, that cells generate a pressure in order to expand in volume.
In turn, the pressure exerted onto the tissue from the environment slows down this volume expansion.
At the \textit{homeostatic pressure}, growth is slowed down to the point where it equals the apoptosis rate: A steady state with constant cell turnover emerges.
However Trepat et al.~\cite{trepat_physical_2009} showed that expanding cellular monolayers were not under pressure, but tensile stress.
This tensile stress was atributed to cellular motility.
Indeed, the two particle growth~(2PG) model~\cite{basan_dissipative_2011} extended by a velocity dependent activation and deactivation of a motility force was remarkably well able to explain the tensile growth~\cite{basan_alignment_2013}.
The velocity dependent activation and deactivation of the motility force leads to an effective alignment interaction between the cell polarity and velocity, orienting the polarity outward, and thus generating tension. 
The model furthermore reproduced swirls in the bulk as found experimentally in confluent monolayers of Madin-Darby Canine Kidney~(MDCK) cells~\cite{angelini_cell_2010,angelini_glass-like_2011}, and fingers at the advancing front reported for wound healling assays~\cite{poujade_collective_2007,ravasio_gap_2015,reffay_interplay_2014,omelchenko_rho-dependent_2003}. 
In 2021, Sarkar et al.~\cite{sarkar_minimal_2021} showed, that alignment interaction are not necessary to explain these.
If the motility force just randomly reorients (Active Brownian Particle -- ABP~\cite{elgeti_physics_2015,fily_athermal_2012,redner_structure_2013}), and the cell-cell adhesion allows for enough wiggle room, cells naturally sort with their polarity pointing outward due to the confinement~\cite{elgeti_wall_2013,fily_dynamics_2014} and thus generating tension. This work however, did not include growth.
We thus combined the 2PG model with the adhesion and motility model of Sarkar et al.
In short, cells consist of two point particles that repell each other by a constant force of magnitude $G$. Upon reaching a critical distance, the cell divides.
Cells interact with an extended Lennard Jones potential of detph $\varepsilon$ and plateau width $\bar\sigma$, experience substrate friction of strength $-\gamma \boldsymbol{v}$, and posess a polarity $\boldsymbol{p}$ in which they exert a spontaneous motility force $f_a=\gamma v_0$.
As in the active brownian particle model, cells reorient by rotational diffusion. See methods for full details. 
Figure \ref{fig:1} shows the surprising result: Even without motility force, the expanding monolayer develops a clear tensile stress!
Adding motility allows the colony to sustain higher tension, and also leads to fingers at the front.
With the right parameters, our minimal model can perfectly match the experimental results of Trepat et al. 2009~\cite{trepat_physical_2009} (Fig.\ref{fig:1}(e)).
Indeed, the phenomenon is rather robust, such that a rather broad range of parameters leads to very good agreement.
In this work, we show how tension arises from the growth response to pressure, and how its influenced by unregulated motility.
Furthermore, we make strong predictions about the average cellular velocities that can be tested experimentally.

\begin{figure}
	\includegraphics[width=\columnwidth]{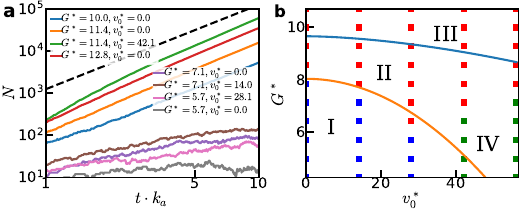}
	\caption{ Phases of growth.
	(a) The number of cells $N$ versus time for some colonies.
	Some weakly growing colonies~(lower four) only grow to a small finite stable size.
	For the colonies that grow to an infinite size~(upper four), $N$ grow quadratic in time in the asymptotic limit, whether it is motile or not.
	The dashed line indicates $N\sim t^2$.
	(b) The phase diagram of colony growth as a function of $G$ and $v_0$.
	Red points grow to an infinite size, blue points grow to finite sizes.
	Green points indicate the scattered phase where the motility is too strong and cause particles to detach and scatter.
	The blue line is the countour line of $P_H=0$ interpolated from simulation measurements.
	The orange line fitted from the simulations separates finite and infinite colonies.
	Between the two lines are the colonies that have negative $P_H$ but grow to an infinite size.
  ($\varepsilon^*=1$ in all simulations)
	}
	\label{fig:1.5}
\end{figure}

Simulations provide insights in the underlying mechanics.
Figure \ref{fig:1} shows the superimposed snapshots of two different colonies from the simulations, as well as their local stress profiles.
The non-motile colony grows as a circular disk, while the motile colony grows with irregular shape with finger-like protrusions.
Both of these colonies are under tension, which builds up from the boundary and is the strongest in the center.
Generally, we can distinguish four different phases of growth (see Fig.\ref{fig:1.5}):
Slow and weakly growing cells form finite steady state colonies (I). 
Above a critical growth strength $G$ or motility $v_0$, the colony grows without reaching a steady state, either under tension (II) or under pressure (III)).
In these cases, the number of cells grows quadratic in time, corresponding to a radial expansion at constant speed, as often observed experimentally~\cite{bru_universal_2003,heinrich_size-dependent_2020,vedula_emerging_2012,trepat_physical_2009}. 
In all these phases, with low or moderate motility, cells do not detach from the colony.
Only above a critical motility, the colony cannot be held together by the interactions anymore and cells behaves like a fast proliferating gas (IV).

\section{Non-motile Quasi-1d Colonies}
To understand how tension arises, one needs to keep in mind that cells have a tendency to proliferate more close to the boundary due to simple mechanical effects~\cite{delarue_mechanical_2013,montel_stress_2011,podewitz_tissue_2015}.
In essence, in order to grow, cells need to deform their surrounding.
Close to the surface, the corresponding strain field is partially cut away, reducing the energetic cost of growth.
On this basis, a quantitative understanding of phases I-III can be optianed from a simple analytical model.
As in Ref.~\cite{delarue_mechanical_2013}, We expand the growth rate $k$ around the homeostatic pressure $P_H$, taking account of the additional growth $\Delta k $ at the surface over a small width $\Delta x$ at boundary $x_0$:
\begin{equation}
	k=\kappa (P_H-P)+\Delta k \Delta x\delta(x-x_0)
\end{equation}
with a response coefficient $\kappa$.
The homeostatic pressure can be negative~\cite{podewitz_tissue_2015}, leading to a negative bulk growth rate at zero pressure.
The surface growth then leads to a stable steady state spheroid in three dimensions with a steady flux of cells from the proliferative rim to the apoptotic core~\cite{podewitz_tissue_2015,delarue_mechanical_2013}.
Friction with the substrate leads to an additional force on the cells, and thus can yield indefinitely growing tensile colonies. 
This mechanism can be best understood in a quasi-one-dimensional setup:
The simulation box is chosen to be very large in $x$ direction and periodic in a short $y$ direction.
The $y$ direction is short enough (about 10 cells) so it can be easily filled, but also large enough that cells can pass each other and form a continuous mass. 
The continuity equation than reads
			\begin{equation}
				\label{eqn:quasi1dcontinuity}
				\partial_t\rho+\partial_x(\rho v)=k\rho.\\ 
			\end{equation}
Assuming constant cell density $\rho$, Eq.(\ref{eqn:quasi1dcontinuity}) becomes 
			\begin{equation}
				\label{eqn:quasi1dcontinuityconst}
				\partial_x v=k.
			\end{equation}

The simulations indicate, that the colony is homeostatically balanced in the $y$ direction~($\sigma_{yy}=-P_H$). With $P=-(\sigma_{xx}+\sigma_{yy})/2$ and defining $P_x=-\sigma_{xx}$ we get
			\begin{equation}
				\label{eqn:quasi1dkandp}
				k = \frac{\kappa}{2} (P_H-P_x) +\Delta k \Delta x\delta(x-x_0). \\ 
			\end{equation}
Force balance in $x$ reads
			\begin{equation}
				\label{eqn:quasi1dforcebalance}
				\partial_x\sigma_{xx}+f_{ext}=0,\\
			\end{equation}
where $f_{ext}$ is the external force density.
Without motility, the only external force on the monolayer is the background friction $f_{ext}=-2\rho\gamma v$.
Combining Eq.(\ref{eqn:quasi1dcontinuityconst}), (\ref{eqn:quasi1dkandp}), and (\ref{eqn:quasi1dforcebalance}) yields
			\begin{equation}
				\label{eqn:quasi1dequation}
				\partial^2_xP_x=\frac{1}{\lambda^2} (P_x-P_H),\\
			\end{equation}
where $\lambda^2=(\rho\kappa\gamma)^{-1}$ (compare Ref.~\cite{podewitz_interface_2016}).
Because pressure is continuous, the boundary condition for the pressure reads $P_x(x_0)=P_x(-x_0)=0$.
Solving Eq.(\ref{eqn:quasi1dequation}) yields
			\begin{equation}
				P_x=P_H(1-\frac{\cosh(x/\lambda)}{\cosh(x_0/\lambda)}).\\
			\end{equation}
For $x_0/\lambda\gg 1$ and $x>0$ we get 
			\begin{equation}
				\label{eqn:quasi1dsolution}
				P_x=P_H(1-e^{\frac{x-x_0}{\lambda}}),
			\end{equation}
i.e. the pressure  builds up from the boundary into the bulk over a length scale $\lambda$, and reaches the homeostatic pressure $P_H$ deep in the bulk.
The velocity profile is obtained from Eq.(\ref{eqn:quasi1dforcebalance}) and Eq.(\ref{eqn:quasi1dsolution}):
			\begin{equation}
				v=P_H\sqrt{\frac{\kappa}{4\rho\gamma}}e^{\frac{x-x_0}{\lambda}}, x<x_0.\\
			\end{equation}
The velocity thus also decays to zero exponentially.

To calculate whether the tissue expands or shrinks, we integrate the growth rate~(Eq.(\ref{eqn:quasi1dkandp})) over the positive half space:
				\begin{equation}
					\frac{1}{2}\frac{\dd N}{\dd t} =\int_0^{x_0}\int_0^{L_y}\rho k\dd x\dd y=L_y(P_H\sqrt{\frac{\rho\kappa}{4\gamma}}+\rho\Delta k\Delta x).\\
				\end{equation}
The first term is the bulk contribution, which happens over a length scale of $\lambda$ from the boudnary, and the second term is the surface growth contribution.
Thus, the colony expands with a constant speed:
			\begin{equation}
				  \label{eqn:quasi1dexpand}
					v(x_0) =\frac{1}{2L_y\rho}\frac{\dd N}{\dd t}=P_H\sqrt{\frac{\kappa}{4\rho\gamma}}+\Delta k\Delta x,
			\end{equation}
which is independent of the colony size.
Thus, if the homeostatic pressure of the colony is positive, it will always grow to an infinite size at a constant speed under pressure (Phase III).
Below a critical (negative) homeostatic pressure of  $P_H^C=-\Delta k\Delta x \sqrt{\frac{4\rho\gamma}{\kappa}}$, the boundary growth cannot compensate the total death in the bulk, and the colony will only grow to a finite size  of $2\lambda \tanh^{-1}(P_H^C/P_H)$~(Phase I), or shrink at a constant speed if the colony is initially larger.
In between, i.e. when $P_H^C<P_H<0$, the colony will grow to an infinite size under tension (Phase II), where the tensile force is balanced by friction forces of the retrograde flow of cells from the proliferating rim to the center.

\begin{figure}
	\includegraphics[width=\columnwidth]{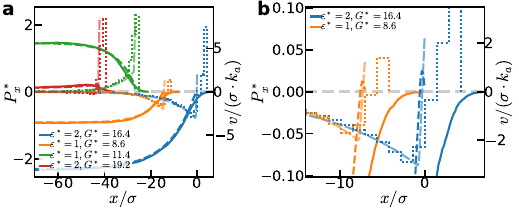}
	\caption{Pressure and velocity profiles of growing non-motile Quasi-1d colonies. 
	The curves of different parameters are shifted along $x$ direction for clarity.
	(a) The pressure profiles~(solid) and velocity profiles~(dotted) obtained from simulations of different parameters and the corresponding theory predictions~(dashed). 
	The pressure builds up from the boudnary to $P_H$ in the bulk exponentially~(except red, whose $P_H$ is positive but smaller than the pressure increase at the boundary due to boundary growth). 
	The velocity profiles decay exponentially in the bulk, and increase drastically but continuously at the boudnary. 
	For colonies with negatiev $P_H$, only the boudnary is moving outward while the rest of the colony is moving towards the center. 
	The theory predictions are calculated from the piecewise solution with all parameters measured independently~(i.e. not fitted, see SI) and $\Delta x$ estimated to be $0.7$. 
	The theory and the simulations show good agreement.
	(b) The pressure profiles of two colonies with negative $P_H$ shown in (a) zoomed in at the front. 
	The theory predicts an increase of pressure at the front for all expanding colonies, even if the homeostatic pressure is negative. 
	However, this pressure is often too small to be observed from the simulations.}
	\label{fig:2}
\end{figure}

Here,we treated surface growth as localized perfectly to the surface in a delta distribution, which leads to a discontinuous jump in the velocity due to Eq.(\ref{eqn:quasi1dcontinuityconst}). 
A more rigorous piecewise solution~(see SI) displays a continuous velocity in the tissue, but otherwise converges to the solution presented above for $\Delta x/\lambda\ll 1$.
\\
Eq.(\ref{eqn:quasi1dexpand}) predicts a constant expansion speed for all colonies in phase II or III and is linear in $P_H-P_H^C$. Indeed, the simulations display a constant expansion speed linear in $G$~(see Fig.S7). We use this constant expansion speed to obtain $\Delta k\Delta x$. 
Obtaining the homeostatic pressure and other bulk tissue properties from bulk simulations~(i.e. without a fit, see SI), and estimating $\Delta x^*=0.7 $ reproduces the simulation data remarkably well~(Fig.\ref{fig:2}). 

For an expanding front, the pressure rises at the boundary, even though the value of this pressure may be too small and narrow to be observed in the simulations.
So for colonies with $P_H<0$, from the boundary to the bulk, the pressure first increases but then peaks, and decreases to $P_H$.
This also applies when the $P_H$ is positive but smaller than the pressure built at the boundary~(See for example the red curve in Fig.\ref{fig:2}) .
In these cases, the velocity profile is negative, i.e. a retrograde flow of cells moving inward, except in a smal region near the boundary.
It is this retrograde flow, that balances the tension inside the colony.
%
Thus our simulations and analytical arguments predict that for non-motile tissues that display tension, cells should exhibit a retrograde flow.  
\\

\section{Motile Quasi-1d colonies}
\begin{figure}
	\includegraphics[width=\columnwidth]{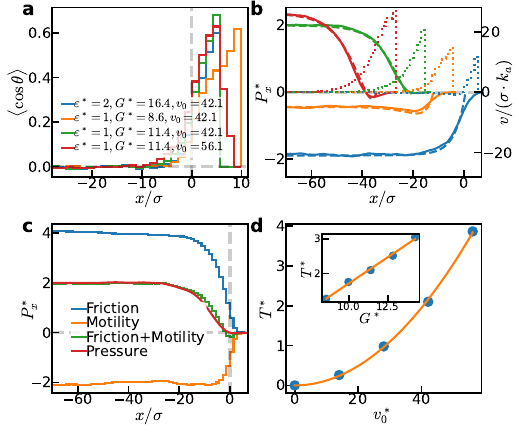}
	\caption{Polarization, pressure, and velocity profiles of motile quasi-1d colonies. 
	(a) The polarization profile of different colonies.
	The polarization has a sharp distribution at the boundary of the colony.
	The maximum polarization is roughly independent of the parameters.
	(b) The pressure and velocity profiles of different motile colonies~(colors as in (a)).
	The curves of different parameters are shifted in $x$ direction for clarity.
	The solid lines are the pressure profiles measured from simulations.
	The dashed lines are calculated from the theory without adjustable parameters.
	The theory matches the simulation results well.
	Motility indeed generates tension at the boundary of the colonies, though the pressure still goes to $P_H$ in the bulk.
	The dotted lines are the velocity profiles measured from simulations.
	(c) Motility force and friction build stress inside the colony.
	They add up to the pressure.
	(d) The tension generated by motility $T$ is quadratic in $v_0$ and linear in $G$ (inset).
	}
	\label{fig:4}
\end{figure}

Without motility, a tensile homeostatic stress is balanced by friction due to a flux of cells inwards from the proliferating boundary. 
Motility adds a second external force to the force balance equation, as cells exert their motility force $f_a=\gamma v_0$ in direction $\boldsymbol{p}=(\cos\theta,\sin\theta)$ on the substrate.\footnote{Note that freely moving cells exert no net force on the substrate, as their motility force is exactly balanced by their friction force.}
The force balance equation thus reads 
\begin{equation}
	\partial_x\sigma_{xx}-2\rho\gamma v+2\rho\gamma v_0\langle\cos\theta\rangle =0,
	\label{eqn:quasi1dforcebalancemotile}
\end{equation}
with the average polarization $\langle \cos\theta\rangle$.
Fig.\ref{fig:4}(a) shows, that in our simulations the polarization is zero in the bulk, but increases sharply towards the boundary, similar to non-proliferating motile colonies~\cite{sarkar_minimal_2021}.
To model the motile colonies, we assume the distribution of the motility force density to be an exponential function of $x$: $2\rho\gamma v_0\langle\cos\theta\rangle=Fe^{\frac{x-x_0}{\lambda_m}}$.
By following similar procedures as above, we obtain the pressure profile for a quasi-1d motile colony:
\begin{equation}
	P_x=P_H(1-e^{\frac{x-x_0}{\lambda}})+T\frac{\lambda^2}{\lambda^2-\lambda_m^2}(e^{\frac{x-x_0}{\lambda_m}}-e^{\frac{x-x_0}{\lambda}}),
	\label{eqn:quasi1dsolutionmotile}
\end{equation}
where $T=\int Fe^{\frac{x-x_0}{\lambda_m}} \dd x = F\lambda_m>0$ is the total tension generated by the motility force over a lengthscale $\lambda_m$.
Motility generates tension at the boundary, but the pressure still builds towards $P_H$ in the bulk over a length scale of $\lambda$ (in the simulations $P_H$ and $\lambda$ depend on $v_0$).
Fig.\ref{fig:4}(b) compares the measured pressure profiles to the analytical expression without adjustable parameters (parameters determined from the orientation profile and independent simulations, see SI), and also displays the measured velocity profiles for different colonies.
Note that for motile colonies the interface roughens, leading to less acurate agreement. 

Comparing simulations with different motility reveals two effects:
On the one hand, motility also favors bulk growth, thus increasing the homeostatic pressure.
Indeed, we find that the homeostatic pressure increases nearly quadratically with motility force~(Fig.S1). 

On the other hand, 
motility generates tension $T$ at the leading edge as predicted by Eq.(\ref{eqn:quasi1dforcebalancemotile}).
This tension can dominate the pressure profile.
Even for positive homeostatic pressure, we observe a dip into tension close to the edge, and for small negative homeostatic pressure, the tension overshoots before relaxing back to the homeostatic one.
The larger the motility force, the stronger this effect.
Importantly, this motility also effects the velocity profile, masking, or even inverting the retrograde flow observed for non-motile tensile colonies.

The simulations allow us to seperate the different contributions to the pressure in the colony.
By integrating the two traction contributions (friction and motility) seperately, we obtain the tension $T$ generated by motility, and the pressure build up due to friction forces.
Consistently, they add up to the total pressure measured via the virial.
Fig.\ref{fig:4}(c) shows these contributions for one exemplary case. 
Fig.\ref{fig:4}(d) and its inset shows the relationship between the total tension generated by motility $T$ as a function of motility $v_0$ and growth force $G$.
The total tension is observed to be quadratic in $v_0$ and linear in $G$. 

To see how the motility induced tension supports tensile colonies, we obtain the expansion speed by integrating the growth rate as above:
\begin{align}
		\label{eqn:quasi1dexpandmotile}
		v(x_0) &=\frac{1}{2\rho\gamma}\left( \frac{P_H}{\lambda}+\frac{T}{\lambda+\lambda_m} \right) +\Delta k\Delta x\nonumber\\ 
		&=\left(P_H+T\frac{1}{1+\frac{\lambda_m}{\lambda}}\right)\sqrt{\frac{\kappa}{4\rho\gamma}}+\Delta k\Delta x.
\end{align}

\begin{figure}
	\includegraphics[width=\columnwidth]{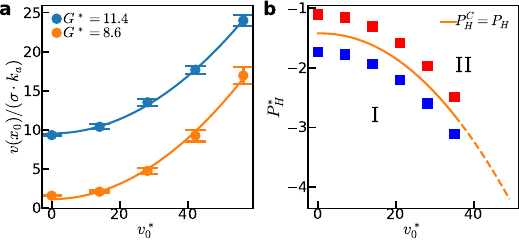}
	\caption{Effects of motility on the growth of motile quasi-1d colonies. 
	(a) The colony expansion speed $v(x_0)$ of colonies with $G^*=11.4$ and $G^*=8.6$ as a function of the motility $v_0$.
	The fit shows that the expansion speed is quadratic in $v_0$.
	Error bars indicate standard deviations.
	(b) The maximum homeostatic tension~($-P^*_H$) at which the colony can still grow to an infinite size~(orange).
	The squares correspond to simulated colonies that grow to an infinite size~(red) and to a finite size~(blue).
	Beyond certain $v_0$, the colony becomes scattered~(dashed).	
	}
	\label{fig:5}
\end{figure}

The expansion speed is still a constant, but with an additional contribution due to motility induced tension.
Fig.\ref{fig:5}(a) shows the simulation data of the expansion speed vs motility $v_0$~(also see Fig.S10).
The expansion speed is quadratic in $v_0$ as expected from the $v_0\rightarrow-v_0$ symmetry of our model.
According to Eq.(\ref{eqn:quasi1dexpandmotile}), the critical homeostatic pressure becomes
\begin{equation}
		\label{eqn:Phc}
		P_H^C = -\Delta k\Delta x \sqrt{\frac{4\rho\gamma}{\kappa}}-T\frac{1}{1+\frac{\lambda_m}{\lambda}},
\end{equation}
thus expanding the phase of indefinitely extending tensile colonies (Phase II).
This allows the colony to sustain higher tension, as shown in Fig.\ref{fig:5}(b).


\section{Two dimensions - growth on a substrate}
\begin{figure}
	\includegraphics[width=\columnwidth]{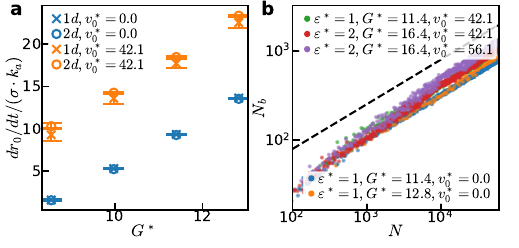}
	\caption{The growth of 2d colonies. 
	(a) The comparison of the expansion speeds of quasi-1d ($dx_0/dt$) and 2d ($dr_0/dt$) colonies.
	They show good agreement, even with motility.
	The error bars indicate standard deviations.
	(b) The relationship between the number of boundary cells $N_b$ and the number of cells $N$.
	The dashed line indicates $N^\frac{1}{2}$. 
	For non-motile colonies, the boundary cell number behaves exactly as $N^\frac{1}{2}$.
	When a motile colony is small, fingers cause fractal-like behaviour.
	But at the limit of large colony, the exponent still goes to $1/2$.
	So the fingers become an undulation of a fixed amplitude, increasing the roughness of the surface.
	The larger the motility, the larger the roughness.
	 }
	\label{fig:6}
\end{figure}

If the shape of the colony does not deviate too much from a circle, the above analysis can be applied to two dimensional growth with radial symmetry~(See SI).
Indeed, the solution for a circular geometry converges to the one dimensional case for radii much larger than $\lambda$~(Fig.\ref{fig:6}(a)).
Importantly however, as shown in Fig.\ref{fig:1}, motility creates fingers at the boundary.
We observe, that a finger is caused by the boundary accumlatation of outward-polarized particles, and that fingering is stronger if the equivalent colony without motility grows slower.
Furthermore, fingers increase the surface area and thus the growth due to boundary growth.
This is partially enhanced by the effect that daughter cells inherit the motility polarizations of their mother cell resulting in a positive feedback loop.
Without this polarization heritage, fingering is reduced and the colony grows slower~(see Supplementary Video 3). 
Given this fingering tendency, one might expect fractal growth of the colonies.
However, as  Fig.\ref{fig:6}(b) shows, while the number of boundary cells is increased, it still scales asymptotically with the square root of the total number of cells - indicating against fractal properties.
Despite all this, Fig.\ref{fig:6}(a) shows that the expansion speeds of motile 2d and quasi-1d colonies are very close, evidencing that the effect of fingers on the overall growth are minor.

\section{Conclusions}
In this paper, we explore the mechanics of growth of cell colonies on a substrate.
Importantly, we find four different phases of growth:
In Phase I, the colony is so contractile it can only grow to finite size.
As the homeostatic pressure increases above the critical pressure determined by Eq.(\ref{eqn:Phc}), the colony grows indefinitely, while remaining under tension (Phase II).
The tissue always reaches its homeostatic state in the center, thus becomming under pressure, once the homeostatic pressure turns positive (Phase III).
Finally, for very strong motility, groups of cells detach, and we arrive at a growing gas (Phase IV). 
The ability to grow indefinitely while under tensile stress is enabeled by two factors: 1. the propensity of cells to grow faster at the interface, and 2. outward directed cellular motility.
These two factors can act independently, or act in consort to balance an even greater tensile core. However, in our simulations the additional growth at the surface is alway present, as it arises from mechanical principles.
Using continuum therory, we quantitatively predict the transition betweeen Phase I,II, and III.
The critical pressure consequentially has two contributions: First, excess growth at the interface results in an retrograte flow of cells, which, due to friction, can balance out a tensile core.
Second, motility in combination with self-generated outward polarization of cells results in tension buildup that can additionally balance a tensile core.

Our findings can help interpret experimental findings of tensile colony growth~\cite{trepat_physical_2009,tambe_collective_2011,sunyer_collective_2016,vedula_emerging_2012}.
Combining these traction force experiments with measuring cellular velocites (for example via particle imaging velocimetry~\cite{angelini_cell_2010,angelini_glass-like_2011,kim_propulsion_2013,vedula_emerging_2012,marel_alignment_2014}) may help to gain further insight into the underlying mechanics.
Especially, the presence of a retrograde flow of cells would be very interesting, even though our model suggests it could be absent due to cellular motility.

Our model has some important shortcommings however.
For one, it does not consider any form of motility alignment, even though biological cells certainly do not reorient randomly~\cite{kim_propulsion_2013,tambe_collective_2011,zisis_disentangling_2022,alert_physical_2020,vercruysse_geometry-driven_2024}.
While usefull to uncover fundamental principles about how tension may arrise, a quantitative matching of experimental data will certainly require some form of alignment.
Our model was able to quantitatively match the data of Trepat et al.~\cite{trepat_physical_2009}, however also suggests a retrograde flow, which has not been reported.
Adding motility alignment could remove this retrograde flow, and a detailed quantitative comparrison of traction and velocity maps may help uncover the true underlying mechanism.
Similar to Ref.~\cite{vercruysse_geometry-driven_2024}, we want to implement various motility alignement mechansims and compare them to experimental data in future works.

Second, part of the force balancing tension in the colonies center comes from friction.
Thus the type of friction plays a key role.
Here, we assumed simple linear friction ($f=-\gamma v$) while cells can certainly display a more complex behavior like dry friction or even an active response to external force.

Third, we have not explored the finger formation in detail.
We expect that similar to competing tissues, linear stability analysis~\cite{williamson_stability_2018} could shed a light on how these fingers form.
Subsequnt comparisson to simulations can than reveal further insights~\cite{buscher_instability_2020}.

Finally, leader cells, supra-cellular actin cables and their interactions certainly play a role in real MDCK colonies~\cite{reffay_interplay_2014,poujade_collective_2007,ravasio_gap_2015} and possibly other cell lines~\cite{omelchenko_rho-dependent_2003,vedula_mechanics_2015}.
This work can only paint the generic picture of how mechanics of tensile growth can function -- a detailed quantitative comparrison will need to take details of the specific cell line into account.

\begin{acknowledgments}
	This project is supported by the China Scholarship Council~(CSC).
\end{acknowledgments}

\appendix

\section{Simulation method}
We use the 2PG model~\cite{basan_dissipative_2011} for tissue growth.
In the model, each cell consists of two particles with diameter $\sigma$.
The two particles are repelling each other with a constant growth force $G$.
When the distance between the two particles exceeds a threshold $r_c$, the cell divides, and a new particle is placed close to each old particle to form two new cells.
Cell apoptosis is modeled as a constant cell removal rate $k_a$.
We implement a Dissipative Particle Dynamics~(DPD) type thermostat for intracell and intercell particle interactions.
The interaction includes a dissipative force 
\begin{equation}
	\boldsymbol{F}^D_{ij}=-\gamma\omega^D(r_{ij})(\boldsymbol{v}_{ij} \cdot \boldsymbol{\hat{r}}_{ij})\boldsymbol{\hat{r}}_{ij}
\end{equation}
and a random force
\begin{equation}
	\boldsymbol{F}^R_{ij}=\mu\omega^R(r_{ij})\varepsilon_{ij}\boldsymbol{\hat{r}}_{ij}.
\end{equation}
Here, $\boldsymbol{v}_{ij}=\boldsymbol{v}_{j}-\boldsymbol{v}_{i}$, $\varepsilon_{ij}$ is a Gaussian variable with zero mean and unit variance, $\omega^D(r_{ij})$ and $\omega^R(r_{ij})$ are weight functions, $\gamma$ is the friction coefficient, which can be chosen independently for intercell and intracell interaction, and $\mu$ is the strength of the random force.
To fulfill the fluctuation-dissipation theorem, $\mu^2=2\gamma k_BT$ and $\omega^D(r_{ij})=[\omega^R(r_{ij})]^2$ must be satisfied.
For intracell particle interaction, we choose $\omega^D(r_{ij})=1$.
And for intercell partcile interaction, we choose $\omega^D(r_{ij})=(1-r_{ij}/R_{PP})^2$, where $R_{PP}$ is the cutoff radius.

For the motility and interaction, we incorporate the modified ABP model of Sarkar et al. 2021~\cite{sarkar_minimal_2021}.
In this model, particles not belonging to the same cell interact with each other with the extended Lennard-Jones~(LJ) potential:
\begin{equation}
	V_{\text{ELJ}}(r)=
	\begin{dcases}
		4\epsilon[(\frac{\sigma}{r})^{12}-(\frac{\sigma}{r})^6], & r<2^{\frac{1}{6}}\sigma \\
		-\epsilon, & 2^{\frac{1}{6}}\sigma \leq r < 2^{\frac{1}{6}}\sigma + \bar{\sigma} \\
		4\epsilon[(\frac{\sigma}{r-\bar{\sigma}})^{12}-(\frac{\sigma}{r-\bar{\sigma}})^6], & 2^{\frac{1}{6}}\sigma + \bar{\sigma}\leq r
	\end{dcases}
\end{equation}
where $\epsilon$ is the interaction strength, $\sigma$ is the diameter of the particle, and $\bar{\sigma}$ is the width of the extended basin, which we choose to be $0.3\sigma$.
Additionally, each particle is subject to a propelling motility force with a constant magnitude $F^M=\gamma_b v_0$, where $\gamma_b$ is the background friction coefficient.
The direction of the motility force is identical for both particles consitituting the same cell, and undergoes a rotational diffusion 
\begin{equation*}
	\dot{\theta_i}=\sqrt{2D_R}\eta^R_i,
\end{equation*}
where $D_R$ is the rotational diffusion constant and $\eta^R_i$ is again a Gaussian white noise.
After a division, the two daughter cells inherit the motility polarization of the mother cell.
Even though the origin of the rotational diffusion can be athermal since this is an active system, we still set the relationship between $D_R$ and $D_T$ to satisfy the Einstein relation $D_T=k_BT/\gamma_B=D_R\sigma^2/3$.
This model contains neither motility alignment nor leader cell mechanisms.

In summary, the total equation of motion of a particle $i$, with $k$ the other particle of the same cell, is
\begin{equation}
	\resizebox{\columnwidth}{!}{$	m\boldsymbol{\ddot{r}}_i = \boldsymbol{F}^G_{ik}+\boldsymbol{F}^D_{ik}+\boldsymbol{F}^R_{ik}+ \sum_{j\neq i,k}(\boldsymbol{F}^{ELJ}_{ij}+\boldsymbol{F}^D_{ij}+\boldsymbol{F}^R_{ij}) +\boldsymbol{F}^{BD}_i+\boldsymbol{F}^{BR}_i+\boldsymbol{F}^M_i,$}
\end{equation}
where each term on the right hand side means growth force, intracell dissipation, intracell fluctuation, intercell extended LJ interaction, inter cell dissipation, intercell fluctuation, background friction, background fluctuation, and motility force respectively.
Equations of motion are integrated with a velocity-Verlet algorithm with an additional calculation of dissipative forces~(DPP-VV from Ref.~\cite{nikunen_how_2003}).
\\
Physical quantities are reported in reduced units, indicated by an asterisk. We use the diameter of the particles $\sigma$, the cell turnover time $k_a^{-1}$, and the interaction strength of the reference tissue $\varepsilon=1$ as the reference parameters.

\end{document}


\title{Free Growth under Tension}
\date{February 5, 2025}
\author{Chenyun Yao}
\author{Jens Elgeti}
\email{j.elgeti@fz-juelich.de}
\affiliation{Theoretical Physics of Living Matter, Institute for Advanced Simulation, Forschungszentrum Jülich, Jülich, Germany}
\maketitle
\section{Piecewise Solution}
Complementary to the analytic model for one-dimensional colonies without motility in the main text (Eq.(1) to Eq.(10)), we derive here a piece-wise solution avoiding the delta-distribution of growth rate at the surface, i.e. the surface growth is modeled explicitly as an additional growth $\Delta k$ in a finite region $\Delta x$ at the surface.
The continuity equation reads: 
\begin{equation}
    \partial_xv=k
\end{equation}
Force balance: 
\begin{gather}
    \partial_x\sigma_{xx}+f_{ext}=0\\
    \partial_xP=-2\rho\gamma v \label{eqn:getv}\\
    \partial^2_xP=-2\rho\gamma\partial_xv=-2\rho\gamma k
\end{gather}
For the growth rate, we specifically include the boundary growth within the boundary growth region: 
\begin{gather}
    k=
    \begin{cases}
        \frac{\kappa}{2}(P_H-P) & \text{for } x\in [0,x_0-\Delta x]\\
        \frac{\kappa}{2}(P_H-P)+\Delta k & \text{for } x\in ( x_0-\Delta x,x_0]\\
    \end{cases}
\end{gather}
We proceed by combing the equations for the two regions. Integrating twice yields the pressure:
\begin{enumerate}
    \item $x\in [0,x_0-\Delta x]$\\
    $P=P_H+D_1e^{x/\lambda}+D_2e^{-x/\lambda}$\\
    \item $x\in ( x_0-\Delta x,x_0]$\\
    $P=P_H+\frac{2\Delta k}{\kappa}+C_1e^{x/\lambda}+C_2e^{-x/\lambda}$\\
\end{enumerate}
As in the main text, we assume mirror symmetry around $x=0$ and thus only treat the positive half space here. The boundary conditions read:\\
\begin{enumerate}
    \item Zero pressure at the boundary $P(x_0)=0$
    \item Pressure and its derivative are continuous, in particular at $x_0-\Delta x$
    \item Mirror symmetry at $x=0$ implies $\frac{\dd P}{\dd x}=0$ at $x=0$
\end{enumerate}
Using the boundary conditions we fix the constants of integration:
\begin{equation}
    P=
    \begin{cases}
    P_H(1-e^{\frac{x-x_0}{\lambda}})+\frac{2\Delta k}{\kappa}(\cosh\frac{\Delta x}{\lambda}-1)e^{\frac{x-x_0}{\lambda}} & \text{for } x\in [ 0,x_0-\Delta x]\\
    (P_H+\frac{2\Delta k}{\kappa})(1-e^{\frac{x-x_0}{\lambda}})+\frac{2\Delta k}{\kappa}e^{-\frac{\Delta x}{\lambda}}\sinh(\frac{x-x_0}{\lambda})& \text{for } x\in ( x_0-\Delta x,x_0]\\
    \end{cases}
\end{equation}
    Note that $\int \rho k\dd x$ is still constant for sufficiently large colonies~($x_0\gg \lambda$).\\
    The velocity profile can be obtained from Eq.(\ref{eqn:getv}):\\
    \begin{equation}
    v=
    \begin{cases}
    \sqrt{\frac{\kappa}{4\rho\gamma}}[P_H+\frac{2\Delta k}{\kappa}(1-\cosh\frac{\Delta x}{\lambda})]e^{\frac{x-x_0}{\lambda}} & \text{for } x\in [ 0,x_0-\Delta x]\\
    \sqrt{\frac{\kappa}{4\rho\gamma}}[(P_H+\frac{2\Delta k}{\kappa})e^{\frac{x-x_0}{\lambda}}-\frac{2\Delta k}{\kappa}e^{-\frac{\Delta x}{\lambda}}\cosh(\frac{x-x_0}{\lambda})]& \text{for } x\in ( x_0-\Delta x,x_0]\\
    \end{cases}
\end{equation}
In the limit of $\Delta x \ll \lambda$, we recover the expansion speed given in the main text:
\begin{equation}
    v(x_0)=\frac{1}{2\rho\gamma\lambda}[P_H+\frac{2\Delta k}{\kappa}(1-e^{-\frac{\Delta x}{\lambda}})]\rightarrow P_H\sqrt{\frac{\kappa}{4\rho\gamma}}+\Delta k\Delta x.\\
\end{equation}
\\

\section{2d Analytical Model}

For a 2d non-motile colony, we assume that it grows as a circular disk. 
The stress tensor in polar coordinates reads
			\begin{equation*}
				\boldsymbol{\sigma}=\begin{pmatrix} \sigma_{rr} & 0 \\ 0 & \sigma_{\theta\theta} \end{pmatrix}, \\
			\end{equation*}
where the off-diagonal terms are zero because of rotational symmetry.
However, in 2d, force balance $\nabla \cdot \boldsymbol{\sigma}+\boldsymbol{f}_{ext}=0$ only gives one useful equation
			\begin{equation}
				\label{eqn:2dequation}
				\partial_r\sigma_{rr}+\frac{1}{r}(\sigma_{rr}-\sigma_{\theta\theta})=2\rho\gamma v_r. \\ 
			\end{equation}
Inspired by the one-dimensional results, we assume $\sigma_{\theta\theta}=-P_H$ and define $P_r \equiv -\sigma_{rr}$ and get $k=\frac{\kappa}{2}(P_H-P_r)$.
We do not consider surface tension and the laplace pressure as in the long time limit the laplace pressure should vanish and we did not observe any obvious contributions from it.
Solving Eq.\ref{eqn:2dequation} with boundary condition $P_r(r_0)=0$ yields
			\begin{equation}
				\label{eqn:2dsolution}
				P_r=P_H(1-\frac{i_0(r/\lambda)}{i_0(r_0/\lambda)}),\\
			\end{equation}
where $r_0\equiv\sqrt{\frac{N}{\pi\rho}}$ is the radius of the colony, $i_0$ is the modified spherical Bessel function of the first kind $\sinh(x)/x$, and $\lambda^2=(\rho\kappa\gamma)^{-1}$.
In the limit of $r_0/\lambda\gg 1$, $P_r \rightarrow P_H(1-e^{(r-r_0)/\lambda})$, which is exactly the same as the quasi-1d solution.
This result matches the pressure profile measured from simulations~(Fig.\ref{fig:4.5}) very well.
The total growth in the bulk is
			\begin{align}
          \int\rho k \dd S &=P_H\sqrt{\frac{\rho\kappa}{4\gamma}}2\pi r_0\frac{\cosh (r_0/\lambda)-1}{\sinh (r_0/\lambda)}\nonumber\\
				&\xrightarrow{r_0/\lambda\gg 1} P_H\sqrt{\frac{\rho\kappa}{4\gamma}}2\pi r_0.
			\end{align}

The resultant cell number growth is thus
\begin{align}
    \frac{\dd N}{\dd t} & =(P_H\sqrt{\frac{\rho\kappa}{4\gamma}}+\rho\Delta k\Delta x)2\pi r_0\\
    \implies \frac{\dd r_0}{\dd t} & =P_H\sqrt{\frac{\kappa}{4\rho\gamma}}+\Delta k\Delta x,
\end{align}
which is the same as $dx_0/dt$ in quasi-1d.
The total growth is proportional to the perimeter of the colony instead of the bulk, thus the reasoning for the quasi-1d setup can also be made for a 2d colony.
Since $dN/dt\sim\sqrt{N}$, $N$ will grow as $t^2$, while $r_0$ will grow linearly in time.
The comparison between the expansion speed of quasi-1d and 2d colonies~(Main text Fig.7(a)) reaveals that they are indeed the same.\\

\section{Stress Measurement in Simulations}
\label{ss}
To measure the local stress of colonies, we divide the the simulation box into a rectilinear grid and measure the stress in each rectangle with
\begin{gather}
    \sigma_{\alpha\beta}=-\frac{1}{A}\left( \sum_i \frac{m_i v_\beta^i \Delta r_\alpha^i}{\Delta t} + \sum_{i,j} r_\alpha^{ij} F_\beta^{ij}   \right),
    \label{eqn:1}
\end{gather}
where $A$ is the area of the rectangle, $m_i$ is the mass of particle $i$, $v_\beta^i$ is the velocity of particle $i$ in $\beta$ direction, $\Delta r_\alpha^i$ is the displacement of particle $i$ in $\alpha$ direction inside the rectangle within a short time $\Delta t$, $r_\alpha^{ij}$ is the position difference between particle $i$ and $j$~(from $i$ to $j$) in $\alpha$ direction inside the rectangle, and $F_\beta^{ij}$ is the force from particle $i$ to $j$ in $\beta$ direction.
\\
We calculate the average pressure profile~(and other profiles) of an expanding quasi-1d colony in the comoving frame of the colony front.
We determine the front position of the colony at a certain time by dividing the simulation box into several channels, find the frontmost particle position in each channel, and take the average.
Then, for all the time frames, we shift the measured profiles to make the positions of the fronts collapse onto the same position, and take the average of the shifted profiles.
\\
For the measurement of the homeostatic pressure, we let the colony grow and fill a square periodic box, and calculate the pressure of the full box with Eq.\ref{eqn:1} and take $P_H=-\text{Tr}(\sigma)/2$.
\\
For quasi-1d colonies, we can also utilize the force balance equation $ \partial_x\sigma_{xx}+f_{ext}=0 $ and integrate the external force density $f_{ext}$ to calculate the stress. 
\\
For motile colonies, there is a debate about whether an active stress contribution should be included in the stress calculation~\cite{das_local_2019}.
However, judging from Fig.4(c) in the main text, our measured pressure matches the pressure integrated from the traction forces well, indicating the addition of the active stress is unnecessary.
\\
For 2d colonies, the stress in general cannot be directly obtained from the integration of traction forces. 
Explicitly, for a 2d circular non-motile colony, the force balance equation is
\begin{equation}
    \partial_r\sigma_{rr}+\frac{1}{r}(\sigma_{rr}-\sigma_{\theta\theta})=2\rho\gamma v_r. 
\end{equation}
However, the additional term, $\frac{1}{r}(\sigma_{rr}-\sigma_{\theta\theta})$, is only non-zero near the boundary, where $r$ is very large.
Thus, we can ignore this additional term and calculate the pressure profile of a large colony by integrating $2\rho\gamma v_r$.
The result for a large 2d non-motile colony is shown in Fig.(\ref{fig:4.5}).
\begin{figure}
    \centering
	\includegraphics[scale=0.4]{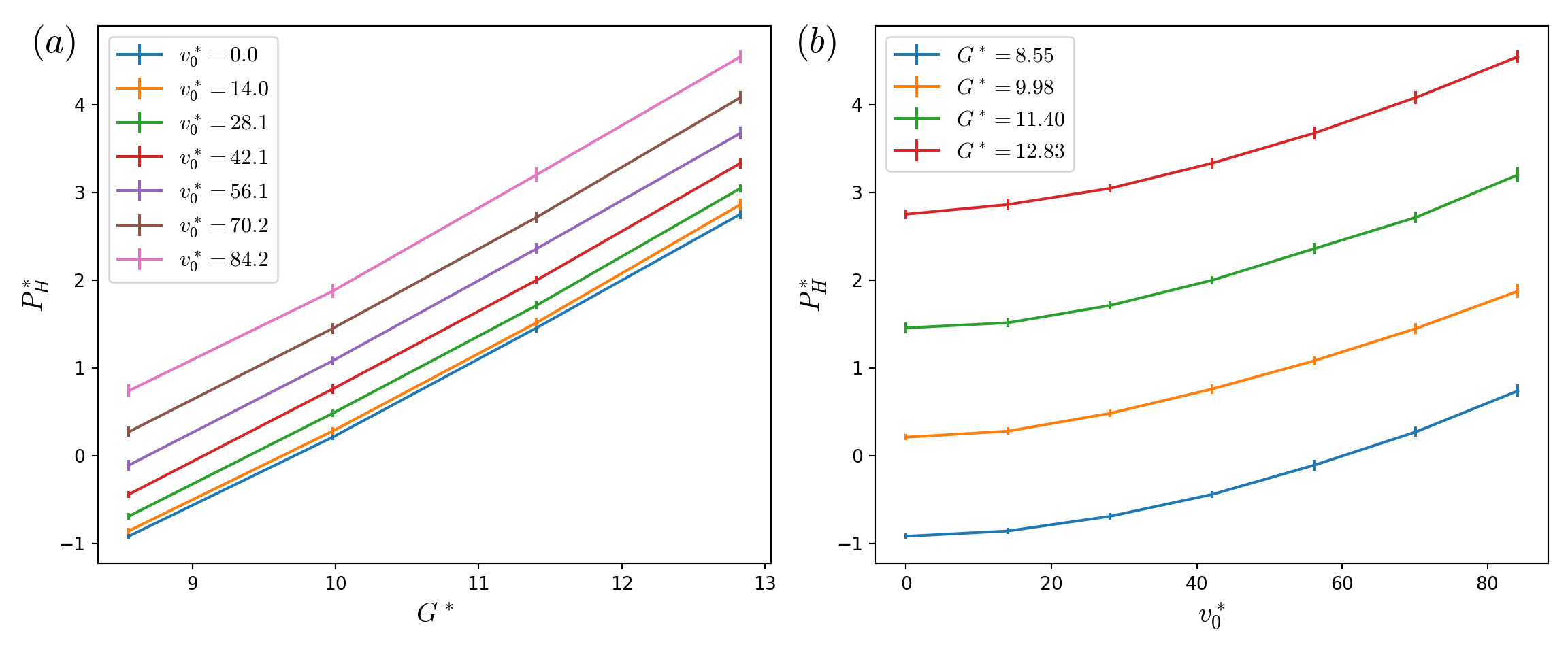}
	\caption{Homeostatic pressure $P_H$ is linear in the growth force $G$ and nonlinear in $v_0$.
    They indicate the affects of $v_0$ and $G$ on $P_H$ are independent.
	}
	\label{fig:1}
\end{figure}

\begin{figure}
    \centering
	\includegraphics[scale=0.5]{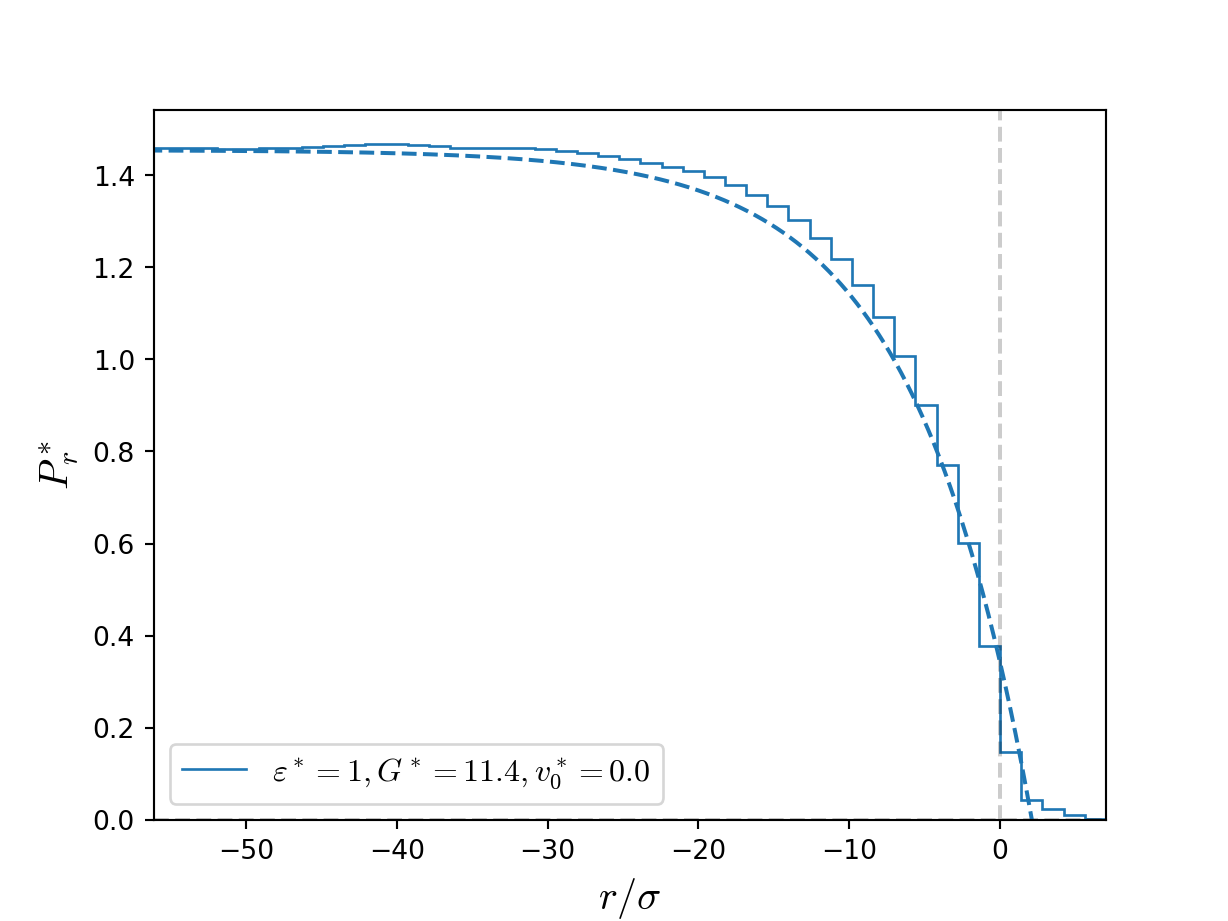}
    \caption{The solid line is the pressure profile $P_r \equiv -\sigma_{rr}$ obtained from the integration of radial traction forces of a large 2d non-motile colony.
    The method is explained in \ref{ss}.
  The dashed curve is calculated from theory without adjustable parameters.
  They show good agreement.
	}
	\label{fig:4.5}
\end{figure}

\section{$\kappa$ Measurement}
\begin{figure}
    \centering
	\includegraphics[scale=0.5]{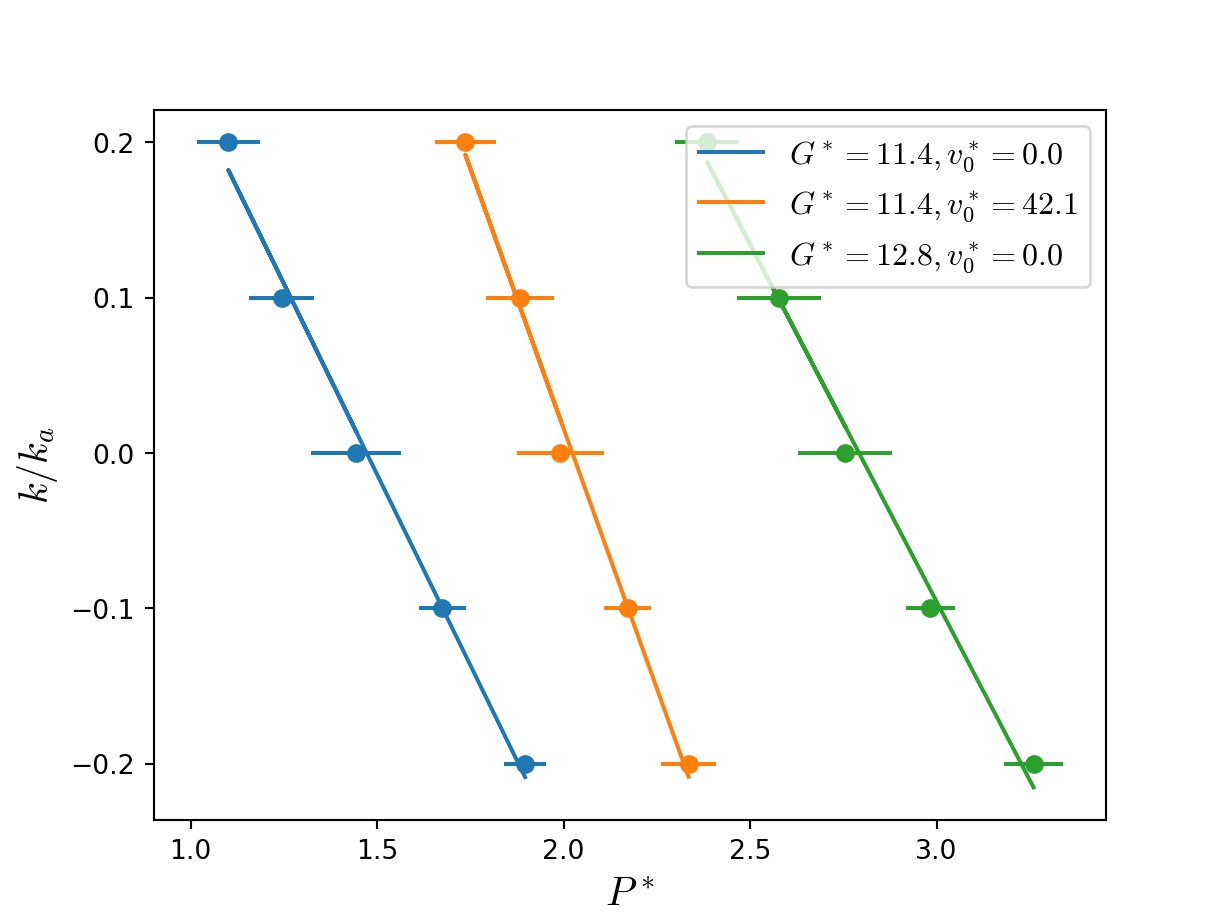}
	\caption{The $k-P$ relationships of different colonies obtained by rescaling the simulation box for $\kappa$ measurement.
  Error bars are standard deviations.
  The slopes are $-\kappa$.
	}
	\label{fig:2.7}
\end{figure}

\begin{figure}
    \centering
	\includegraphics[scale=0.5]{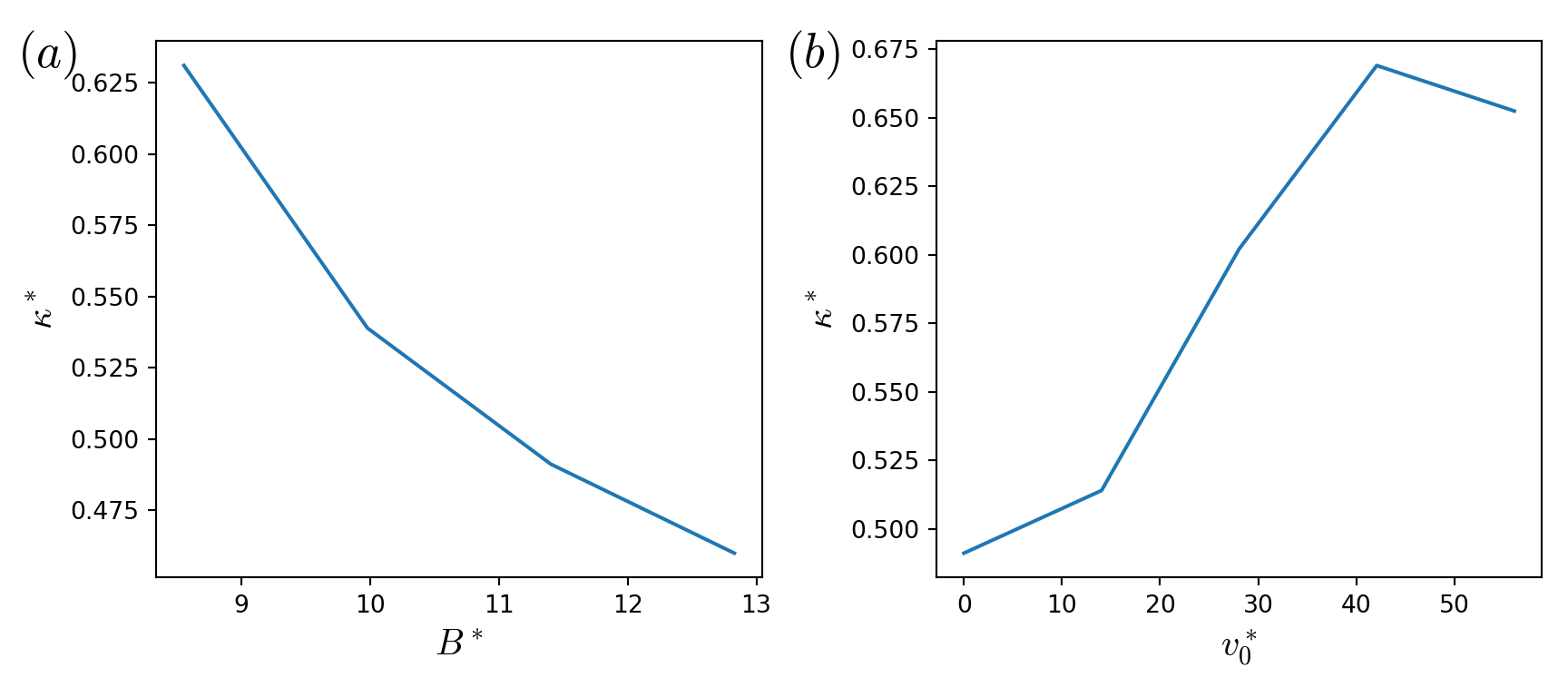}
    \caption{The dependence of $\kappa$ on $G$ with $v_0^*=0$~(a) and $v_0$ with $G^*=11.4$~(b).
	}
	\label{fig:3}
\end{figure}

To measure the pressure response coefficient $\kappa$, we first fill the simulation box with the tissue to be measured.
At every time unit, we rescale the simulation box and the positions of all cells accrding to an imposed rate while measuring the pressure of the simulation box.
If the rate is small enough, the tissue will keep up the expansion/shrinkage of the simulation box, so that the tissue is growing/shrinking at the chosen rate.
We can select different rates and obtain $\kappa$ from the fitting of $k=\kappa(P_H-P)$~(Fig.\ref{fig:2.7}).
The results are shown in Fig.\ref{fig:3}.
\\

\section{Determining Boundary Cells}
\begin{figure}
    \centering
	\includegraphics[scale=0.5]{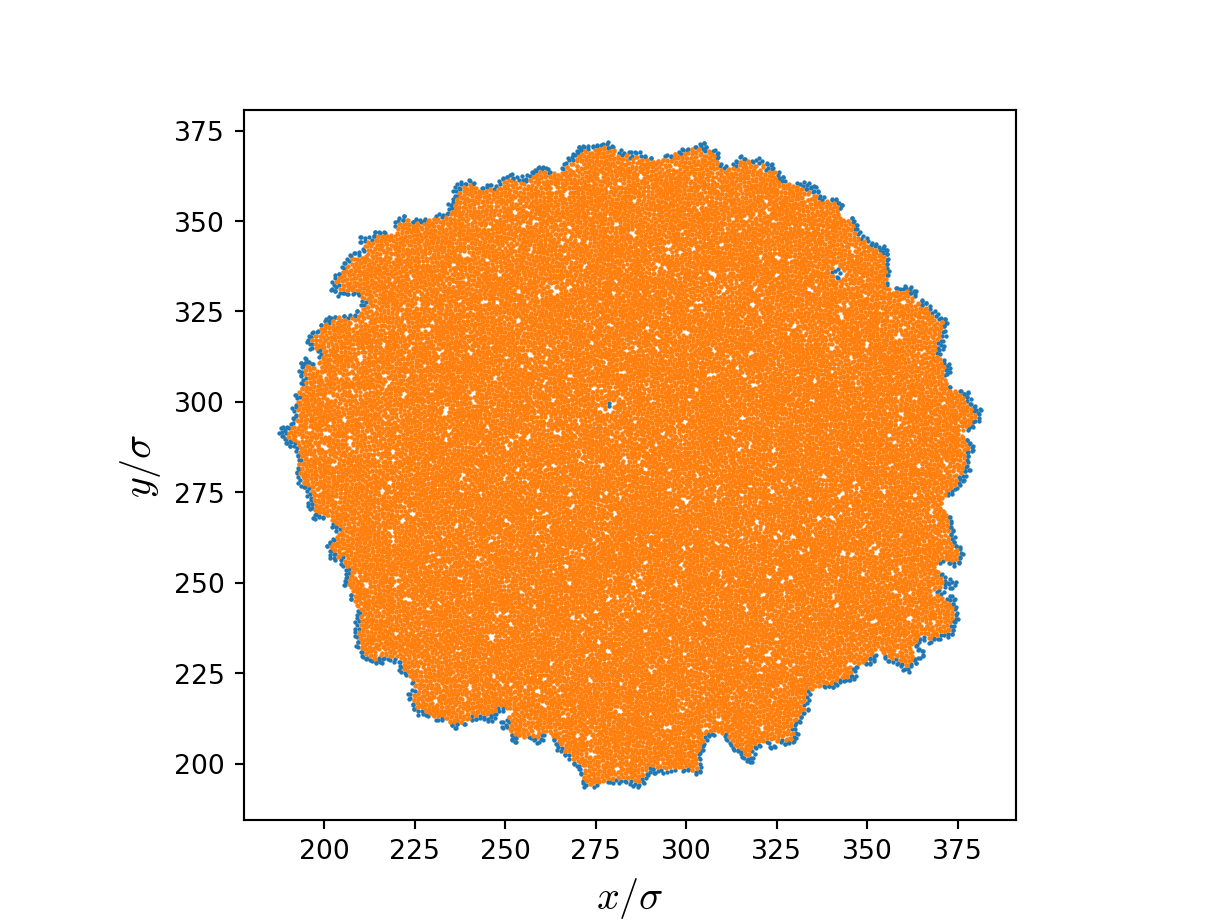}
    \caption{A snapshot of a 2d colony with the boundary cells selected by the k-d tree method colored in blue.
    Some boundary cells at the concave parts of the boundary are not selected.
	}
	\label{fig:5}
\end{figure}

In Fig.6(b) of the main text, we plotted the number of boundary cells vs the number of cells to test for a possible fractal boudnary.
The boundary cells of a colony are determined by the number of neighbour cells of a cell.
We use the k-d tree method~\cite{bentley_multidimensional_1975} to determine the number of neighbours of each cell, then set the threshold as a portion of the mean neighbour count.
All the cells with neighbours less than the threshold are marked as boundary cells.
With suitable choices of the neighbour counting range and the threshold, we can select exactly one layer of the boundary cells~(Fig.\ref{fig:5}).

\section{Additional Results}
\begin{figure}
    \centering
	\includegraphics[scale=0.5]{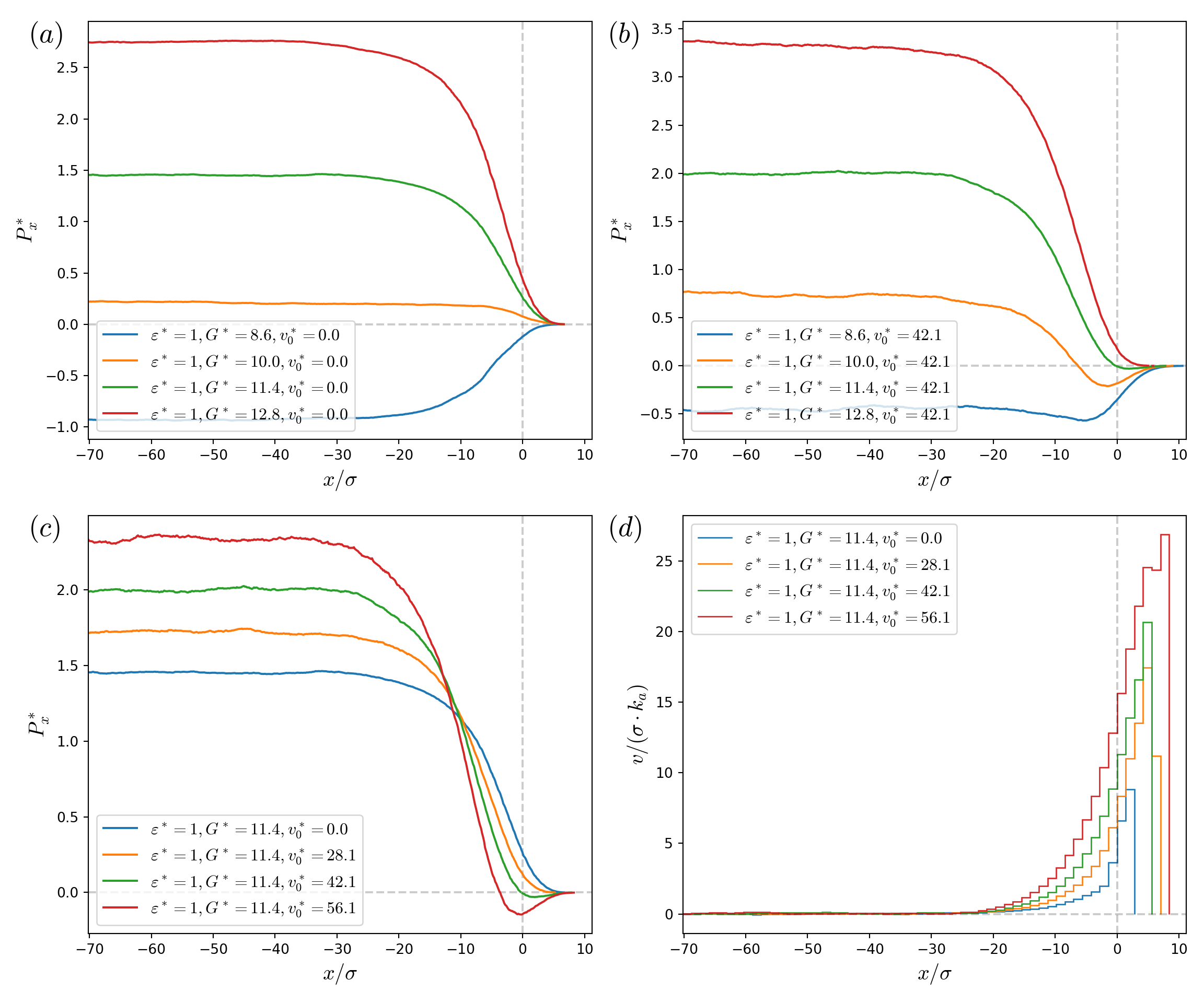}
	\caption{Comparison of pressure and velocity profiles between different colonies.
	(a) The pressure profiles of non-motile colonies with different growth force $G$.
	(b) The pressure profiles of motile colonies with different growth force $G$.
	(c) The pressure profiles of colonies with different motility $v_0$.
  (d) The velocity profiles of the colonies in (c).
	}
	\label{fig:4}
\end{figure}

As further illustrations of the effects of motility $v_0$ and growth force $G$ on colony expansion, we present here~(Fig.\ref{fig:4}) further quantifications of the pressure and velocity profiles of various parameters.
\begin{figure}
    \centering
	\includegraphics[scale=0.5]{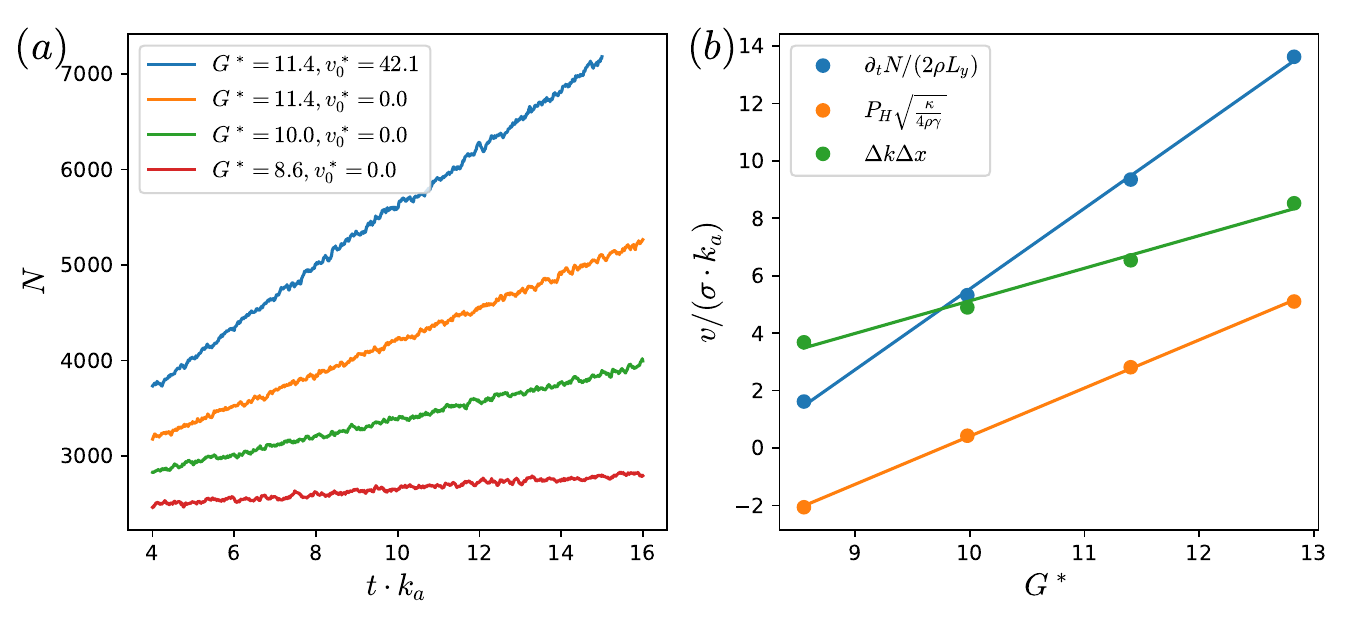}
	\caption{Quasi-1d colony expansion. 
	(a) The numbers of cells of various parameters all grow linear in time. 
	(b) Different contributions of the colony expansion speed $v$ as functions of the growth force $G$. 
	The surface growth contribution (green) is obtained from the difference between the total expansion speed (blue) and the bulk contribution (orange). 
	All of them have linear relationships with $G$. 
	The error of the expansion speed is negligible.}
	\label{fig:main3}
\end{figure}

\begin{figure}
    \centering
	\includegraphics[scale=0.5]{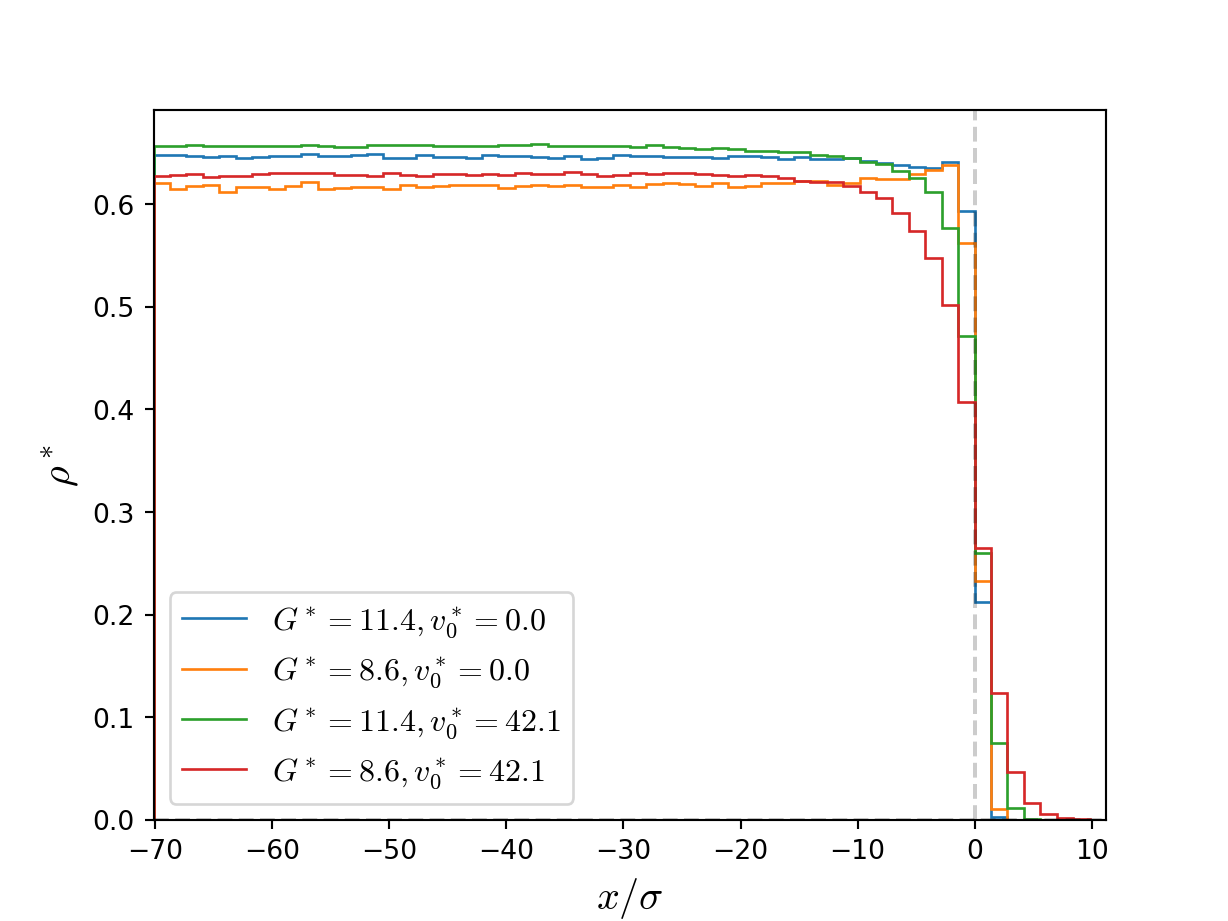}
	\caption{Density profiles of different colonies.
    Density is approximately constant in the bulk.
    The densities of non-motile colonies drop sharply from the bulk density to zero at the boudnary, while the densities of motile colonies drop to zero over a wider range due to fingering.
    }	
	\label{fig:2}
\end{figure}
\begin{figure}
    \centering
	\includegraphics[scale=0.5]{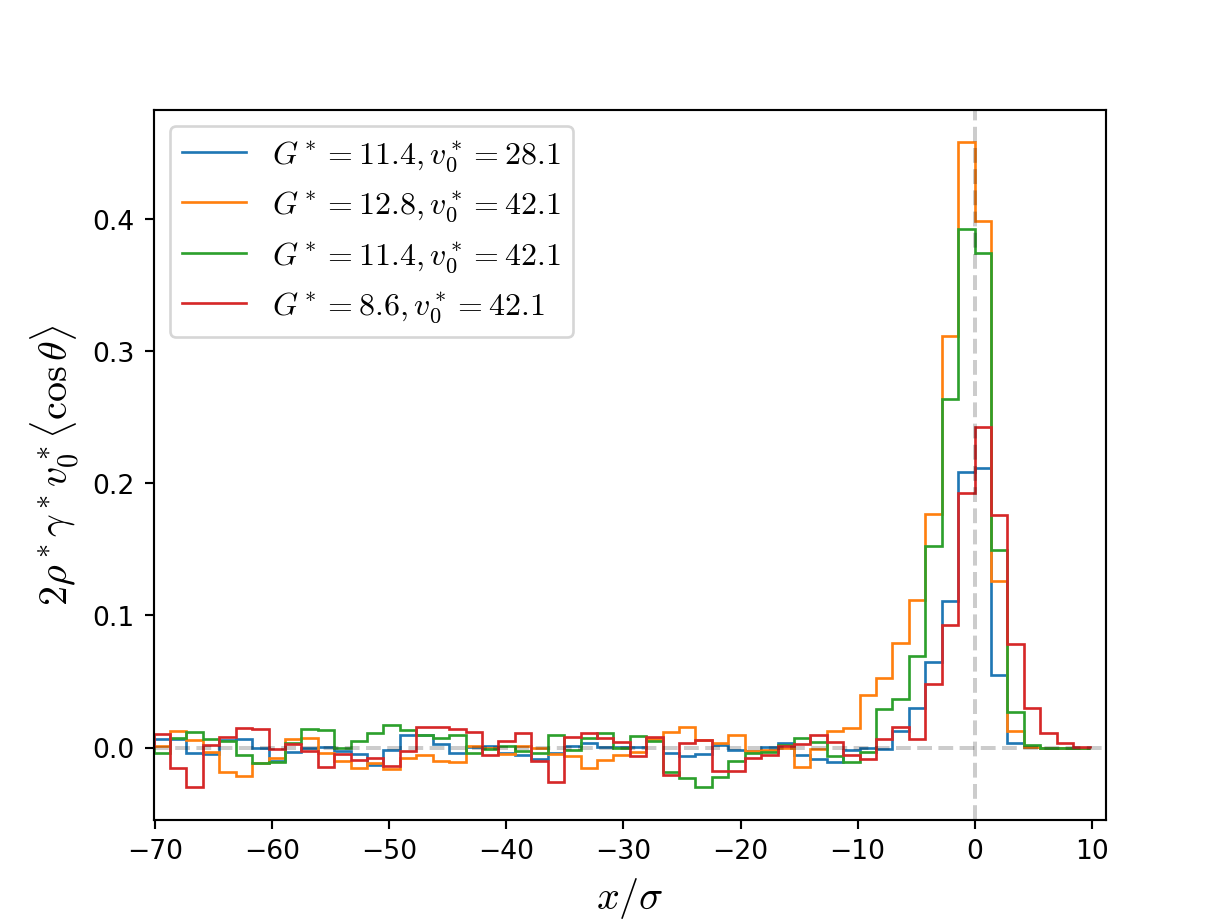}
	\caption{The motility force density profiles of different colonies.
    Due to the fact that the density profile is not a step function for motile colonies, the real motility force density is not an exponential function.
    We estimate $\lambda_m$ from the rising part of the force density profile via $\lambda_m=\frac{\text{tension generated by the rising part}}{\text{maximum of motility force density}}$.
	}
	\label{fig:2.3}
\end{figure}
\begin{figure}
    \centering
	\includegraphics[scale=0.5]{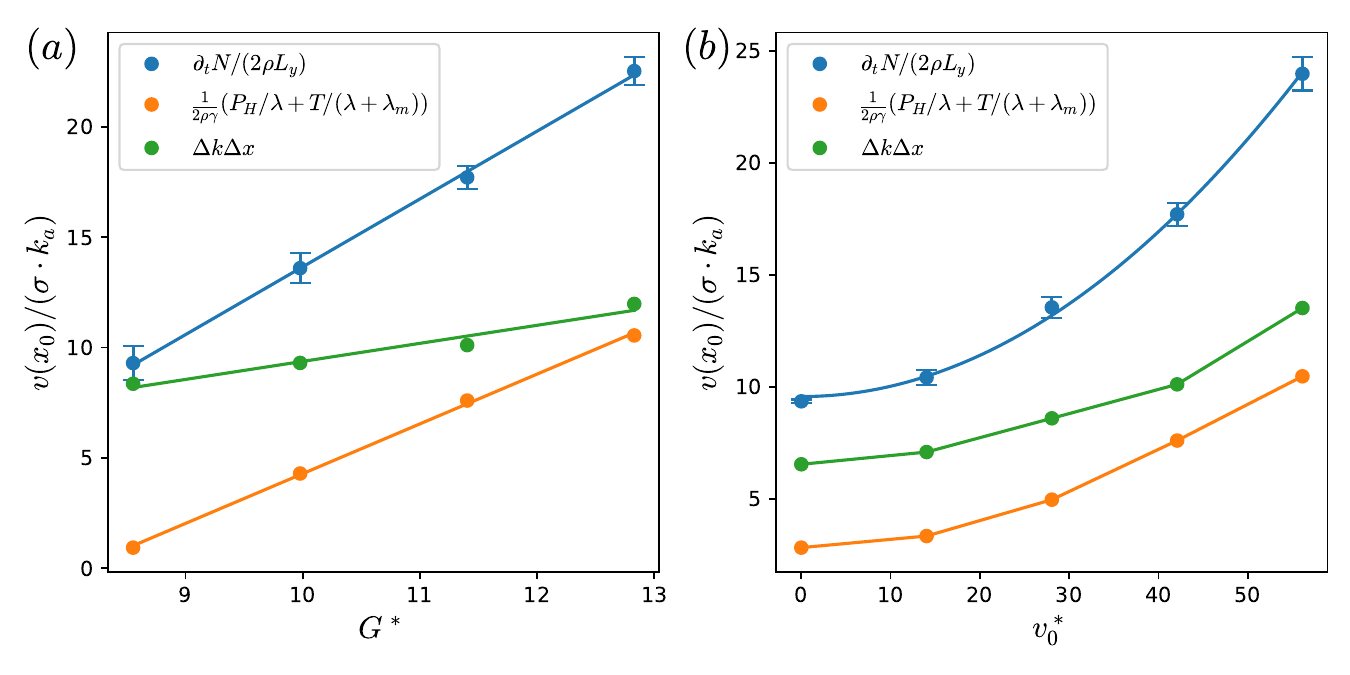}
	\caption{Colony expansion speed of motile Quasi-1d colonies. 
	(a) Different contributions of the colony expansion speed $v$ of colonies with $v_0^*=42.1$ as functions of the growth force $G$.
	The expansion speed and the contributions are still linear in $G$.
	(b) Different contributions of the colony expansion speed $v$ of colonies with $B^*=11.4$ as functions of the motility $v_0$.
	The fit shows that the expansion speed is quadratic in $v_0$.
	The surface growth contribution (green) is obtained from the difference between the total expansion speed (blue) and the bulk contribution (orange). 
	The error bars indicate standard deviations.
	}
	\label{fig:decomposition}
\end{figure}

\vspace{20pt}
\textbf{Supplementary Movie 1} Example growth of a non-motile colony with a negative homeostatic pressure~( $\varepsilon^*=2, G^*=16.4, v_0^*=0$). Left: particle positions. Right: local pressure.

\textbf{Supplementary Movie 2} Example growth of a motile colony with a negative homeostatic pressure~( $\varepsilon^*=2, G^*=15.0, v_0^*=42.1$). Left: particle positions. Right: local pressure.

\textbf{Supplementary Movie 3} Growth of two motile quasi-1d colonies with the same parameters except the cells above inherits polarization after division while the cells below randomly reorients. The former has fingers and expand faster.